\documentclass[aps,prb,twocolumn,superscriptaddress]{revtex4-1}
\usepackage{amsmath,tikz,subcaption,graphicx,todonotes,enumerate,booktabs,multirow}

\DeclareMathOperator{\Tr}{Tr}
\newcommand{\ul}[1]{\underline{#1}}

\begin{document}
\title{An introduction to microscopic theories for inhomogeneous liquids: getting started with density functional theory and the wetting transition}
\author{Adam P.~Hughes}
\affiliation{Department of Mathematical Sciences, Loughborough University, Loughborough, Leicestershire, LE11 3TU, UK}
\author{Uwe Thiele}
\affiliation{Department of Mathematical Sciences, Loughborough University, Loughborough, Leicestershire, LE11 3TU, UK}
\affiliation{Institut f\"ur Theoretische Physik, Westf\"alische
Wilhelms-Universit\"at M\"unster, Wilhelm Klemm Str.\ 9, D-48149 M\"unster, Germany}
\author{Andrew J.~Archer}
\affiliation{Department of Mathematical Sciences, Loughborough University, Loughborough, Leicestershire, LE11 3TU, UK}
\date{\today}

\begin{abstract}
Classical density functional theory (DFT) is a statistical mechanical theory for calculating the density profiles of the molecules in a liquid. It is widely used, for example, to calculate the density distribution of the molecules in the vicinity of a confining wall, the interfacial tension, the wetting behaviour and many other properties of nonuniform liquids. DFT can however be somewhat daunting to students entering the field, because of the many connections to other areas of liquid-state science that are required and used to develop the theories. Here we give an introduction to some of the key ideas, based on a lattice-gas (Ising) model fluid. This builds on knowledge covered in most undergraduate statistical mechanics and thermodynamics courses and so students can quickly get to the stage of calculating density profiles, etc for themselves. We derive a simple DFT for the lattice-gas and present some typical results that can readily be calculated using the theory.
\end{abstract}

\maketitle

\section{Introduction}
The behaviour of liquids at interfaces and in confinement is a fascinating and important area of study. For example, the behaviour of a liquid under confinement between two surfaces determines how good a lubricant that liquid is. The nature of the interactions between the liquid and the surfaces is crucial. Consider, for example, the teflon coating on non-stick cooking pans, used because water does not adhere to (wet) the surface. One can approach the problem from a mesoscopic fluid-mechanical point of view, see for example the excellent book by de Gennes, Brochard-Wyart and Quer\'e.\cite{deGennes04} However, if a microscopic approach is required, which relates the fluid properties at an interface to the nature of the molecular interactions, then one must start from statistical mechanics. There are a number of books such as Refs.~\onlinecite{Henderson92,Rowlinson02,Davis96,Hansen06} which provide a good starting point. All of these include a discussion on classical density functional theory (DFT) which is a theory for determining the density profile of a fluid in the presence of an external potential, such as that exerted by the walls of a container. 

DFT is a statistical mechanical theory, where the aim is to calculate average properties of the system being studied. In statistical mechanics, the central quantity of interest is the partition function $Z$ and once this is calculated, all thermodynamic quantities are given. However, $Z$ a sum over all the possible configurations of the system, can rarely be evaluated exactly. Instead of focussing on $Z$, in DFT we seek to develop good approximations for the free energy. It can be shown that the free energy is a functional of the fluid density profile $\rho(\ul{r})$ and the equilibrium profile is that which minimises the free energy. Over the years, a great many different approximations for the free energy functionals have been developed, generally by making contact with results from other branches of liquid-state physics. There are now quite a few lecture notes and review articles on the subject.\cite{Evans79, Evans92, Lutsko10, Wu07, Wu06, Tarazona08, Lowen10} This rather large literature can make learning about DFT rather daunting. One of us (AJA) has found in teaching this subject that a good place for students to start learning about the properties of inhomogeneous fluids, is by considering a simple lattice gas (Ising) model. This allows students to avoid much of the liquid-state physics and functional calculus that can be daunting for undergraduates when embarking on studying DFT and its applications.\cite{endnote1} The advantage of starting from a lattice-gas model is that one can quickly develop a simple mean-field DFT (described below) and then proceed to calculate the bulk fluid phase diagram and study the interfacial properties of the model, determining the wetting behaviour, finding wetting transitions and all the other interesting phenomenology of liquids at interfaces. The computer algorithms required to solve these equations are fairly simple. Thus, the threshold for entering the subject and getting to the point where a student can calculate things for themselves is much lower via this route, than most other routes that we can think of.

The aim of this paper is two-fold: (i) to derive the mean-field DFT for an inhomogeneous lattice-gas fluid, whilst explaining the physics of the theory. This presentation assumes the reader has had introductory statistical mechanics and thermodynamics courses, but little else beyond that. (ii) To illustrate the types of quantities that DFT can be used to calculate, such as the surface tension of the liquid-gas interface, to study wetting behaviour or to answer the question ``what is the shape of a drop of liquid on a surface?'' We also give some exercises for students.

This paper is laid out as follows: in \S\ref{sec:continuum} we introduce the statistical mechanics of simple liquids. We set up the DFT model in \S\ref{sec:discrete} and \ref{sec:potentials}. The bulk fluid phase diagram is discussed in \S\ref{sec:phase}. We describe the iterative method for solving the model in \S\ref{sec:solving} before displaying some typical results in \S\ref{sec:results}. Finally, some conclusions are drawn in \S\ref{sec:conclusions}.

\section{Statistical Mechanics of Simple Liquids}\label{sec:continuum}
We consider a fluid composed of $N$ atoms/molecules in a container. What follows is also relevant to colloidal suspensions and so we simply refer to the atoms, molecules, colloids, etc as `particles'. The energy is a function of the set of position and momentum coordinates, $\ul{r}^N \equiv \{ \ul{r}_1,\ul{r}_2,\dots , \ul{r}_N \} $ and $\ul{p}^N \equiv \{ \ul{p}_1,\ul{p}_2,\dots , \ul{p}_N \} $ respectively, and is given by the Hamiltonian \cite{Hansen06}
\begin{equation}\label{eq:hamiltonian}
{\cal H}(\ul{r}^N,\ul{p}^N) = K(\ul{p}^N) + E(\ul{r}^N),
\end{equation}
where $K$ is the kinetic energy
\begin{equation}\label{eq:kinetic}
K=\sum_{i=1}^N \frac{\ul{p}_i^2}{2m},
\end{equation}
and $E$ is the potential energy due to the interactions between the particles and also to any external potentials such as those due to the container walls. When treating the system in the canonical ensemble, which has fixed volume $V$, particle number $N$ and temperature $T$, the probability that the system is in a particular state is\cite{Hansen06, Davis96, Plischke06, Reichl09}
\begin{equation}
f(\ul{r}^N,\ul{p}^N)=\frac{1}{h^{3N}N!}\frac{e^{- \beta {\cal H}}}{Z},
\end{equation}
where 
\begin{equation}\label{eq:cPartition}
Z = \frac{1}{h^{3N} N!} \int d \ul{r}^N \int d \ul{p}^N e^{- \beta {\cal H}},
\end{equation}
is the canonical partition function, $h$ is Plank's constant and $\beta = (k_BT)^{-1}$ where $k_B$ is Boltzmann's constant. The partition function allows macroscopic thermodynamic quantities to be related to the microscopic properties of the system which are defined in ${\cal H}$ (see below).

The kinetic energy contribution \eqref{eq:kinetic}, is solely a function of the momenta $\ul{p}^N$, and $E(\ul{r}^N)$, the precise form of which is yet to be defined, only depends on the positions of the particles $\ul{r}^N$. This allows the partition function \eqref{eq:cPartition} to be simplified by performing the Gaussian integrals over the momenta to obtain
\begin{align}
Z & = \frac{1}{h^{3N}} \int d\ul{p}^N e^{- \beta \sum_{i=1}^N \frac{\ul{p}_i^2}{2m}} Q, \nonumber \\
   & = \frac{1}{h^{3N}} \int e^{- \beta \frac{\ul{p}_1^2}{2m}} d\ul{p}_1 \dots \int e^{- \beta \frac{\ul{p}_N^2}{2m}} d\ul{p}_N Q, \nonumber \\
   & = \frac{1}{h^{3N}} \left(\sqrt{\frac{2m \pi}{ \beta}}\right)^{3} \dots \left(\sqrt{\frac{2m \pi}{ \beta}}\right)^{3} Q, \nonumber \\
   & = \left(\sqrt{\frac{2m \pi}{ \beta h^2}}\right)^{3N} Q, \nonumber \\
   & = \Lambda^{-3N}Q,
\end{align}
where $ \Lambda$ is the thermal de Broglie wavelength and 
\begin{equation}\label{eq:configInt}
Q = \frac{1}{N!}\int d\ul{r}^N e^{- \beta E},
\end{equation}
is the configuration integral.\cite{Plischke06} Thus, the partition function is just the configuration integral multiplied by a factor that depends on $N$, $T$ and $m$ and so the value of $\Lambda$ is irrelevant for determining the state of the system. Changing $ \Lambda$ just adds a constant to the free energy per particle [see Eq.~\eqref{eq:helmholtz1}] and so we safely assume $ \Lambda=1$.

Evaluating $Q$ is the central problem here and, in general, this can not be done and so approximations are required. In the following section we develop a simple lattice model approximation that allows progress. Note that the system described above has been analysed in the canonical ensemble. We discuss below how the system can instead be considered in the grand canonical ensemble.

\section{Discrete Model}\label{sec:discrete}
\subsection{Defining a Lattice}

\begin{figure}[t]
\begin{center}
\begin{subfigure}[!]{0.2\textwidth}
\begin{center}
\includegraphics{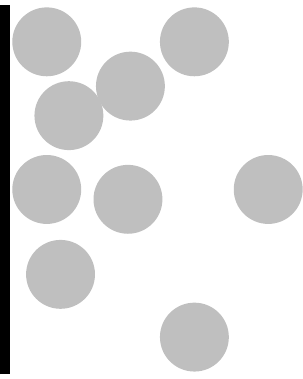}
\caption{}\label{subfig:nolattice}
\end{center}
\end{subfigure}
\begin{subfigure}[!]{0.2\textheight}
\begin{center}
\includegraphics{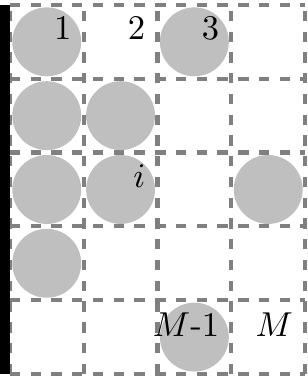}
\caption{}\label{subfig:lattice}
\end{center}
\end{subfigure}
\caption{Illustration of how a free system, (a), may be discretized in space by setting the particles on a lattice, (b).}
\label{fig:lattice}
\end{center}
\end{figure}

We assume that the fluid is two dimensional (2D), to simplify the analysis. However, everything can easily be extended to a three dimensional (3D) system. We imagine a lattice discretises the space occupied by the fluid and so any configuration of particles may be described by a set of lattice occupation numbers $\{n_1,n_2,\dots, n_N\} \equiv \{n_i\}$ which define if the lattice sites are filled ($n_i=1$) or empty ($n_i=0$) with $n_i$ being the occupation number of site $i$. The width of each lattice site is set as $\sigma$, the diameter of a particle, and there are $M$ sites. We set $ \sigma=1$ throughout and use this as our unit of length. The particles are assumed to be spherical, so that their orientation is not important. We now find that the configurational integral in Eq.~\eqref{eq:configInt} becomes a sum over the lattice sites. Note the short hand $i\equiv(k,l)$, where $k$ and $l$ are integer indices defining the 2D lattice.

\subsection{Energy of the System}
To proceed, we must define the potential energy contribution to the Hamiltonian, $E$. We assume the following form
\begin{equation}\label{eq:energy}
E=\sum_{i=1}^M n_iV_i-\sum_{i,j} \epsilon_{ij} n_in_j.
\end{equation}
The first term is the contribution from the external potential $V_i$ and the second term is the energy contribution from pair interactions between particles. We assume that there are no three-body or higher interactions. The interaction energy between particles at two lattice sites $i$ and $j$ is $ \epsilon_{ij}$. This gets smaller as the distance between them increases and so $ \epsilon_{ij}$ has the property that as $|i-j| \rightarrow \infty$, $ \epsilon_{ij} \rightarrow 0$. The term $-\sum_{i,j} \epsilon_{ij} n_in_j $ denotes a sum over all pairs of lattice sites in the system. Considering only pair interactions greatly simplifies the task of evaluating the partition function, but it can still be very arduous to evaluate this sum even for a moderately sized system. The probability of being in a particular configuration, $\{n_i\}$, for a fixed number of particles $N$, is now 
\begin{equation}\label{eq:probability}
P(\{n_i\})= \frac{e^{- \beta E(\{n_i\})}}{Z},
\end{equation}
with the partition function defined as 
\begin{equation}\label{eq:partition}
Z = \sum_{\textrm{all states}} e^{- \beta E_{\textrm{state}}},
\end{equation}
where `state' is a shorthand for a particular allowed set of occupation numbers $\{n_i\}$. Note the relation to the configuration integral in Eq.~\eqref{eq:configInt}, since the sum over all states approximates the continuum integral $(N!)^{-1}\int d\ul{r}^N (\cdots)$.

\subsection{Helmholtz Free Energy}
The Helmholtz free energy is related to the partition function as follows\cite{Hansen06, Plischke06, Reichl09}
\begin{equation}\label{eq:helmholtz1}
F = - k_BT \ln Z.
\end{equation}
All other thermodynamic quantities are obtained from derivatives of $F$. However, we are still unable to evaluate the sum in Eq.~\eqref{eq:partition} and as a consequence can not calculate $F$. Under certain assumptions we can make some progress: consider the system where there is no external field, i.e.\ $V_i=0$,  and that $\epsilon_{ij}=0$, so that the particles do not interact with each other. From Eq.~\eqref{eq:energy}, this gives $E=0$ for all configurations and from Eq.~\eqref{eq:probability} we observe that 
\begin{equation}
P(\{n_i\}) = \frac{1}{Z},
\end{equation}
i.e.\ that all configurations are equally likely. From Eq.~\eqref{eq:partition}, we see that $Z$ is just the number of possible states, which, for a system of $M$ lattice sites containing $N$ particles, is
\begin{equation}
Z= \frac{M!}{N!(M-N)!}.
\end{equation}
For large systems, i.e.\ when both $M$ and $N$ are large, this can be simplified using Stirling's approximation, $\ln(N!) \approx N \ln N - N$, which, with Eq.~\eqref{eq:helmholtz1}, gives
\begin{equation}\label{eq:helmholtz2}
F = -k_BT \left[M\ln M - N\ln N - (M-N)\ln(M-N) \right]. 
\end{equation}
The number density of particles in the system is ${\rho = N/M}$ (recall $\sigma=1$) and so Eq.~\eqref{eq:helmholtz2} gives
\begin{equation}\label{eq:helmholtz3}
F = Mk_BT \left[ \rho \ln \rho + (1- \rho) \ln(1- \rho)\right].
\end{equation}
This homogeneous fluid has a uniform density $\rho$ throughout. However, for an inhomogeneous fluid in the presence of a spatially varying external potential $V_i$ we should expect the density to vary in space. The average density at lattice point $i$ is defined as
\begin{equation}\label{eq:rho_defn}
\rho_i=\langle n_i \rangle,
\end{equation}
i.e.\ it is the average value of the occupation number at site $i$, over all possible configurations: $\langle \cdots \rangle=\sum_{\textrm{all states}} (\cdots) P_{\textrm{state}}$. We now obtain an approximation for the free energy of the inhomogeneous fluid. 

\subsection{The Grand Canonical Ensemble}
We previously treated the system in the canonical ensemble with a fixed $N$, $T$ and volume $V$ (strictly, this is an area since the fluid is 2D but we refer to area as `volume' throughout). Now we consider the system in the grand canonical ensemble with fixed $V$ and $T$ but now $N$ can vary by exchanging particles with a reservoir. The reservoir has a fixed chemical potential $\mu$, and as the system is connected to this reservoir it has the same chemical potential (recall that the chemical potential is the energy required to insert a particle into the system). Physically, the easiest way to conceive the grand canonical ensemble is to imagine the system as being a subsystem of a much larger structure, with which it can exchange particles, and where the reservoir fixes $T$ and $ \mu$ in the subsystem.

The probability of a grand canonical system being in a particular state is [cf.\ Eq.~\eqref{eq:probability}]
\begin{equation}\label{eq:gprob}
P(\{n_i\}) = \frac{e^{- \beta(E - \mu N)}}{ \Xi},
\end{equation}
where the number of particles in the system is 
\begin{equation*}
N = \sum_{i=1}^M n_i,
\end{equation*}
The normalisation factor $\Xi$ is the grand canonical partition function 
\begin{equation}\label{eq:gpartition}
\Xi = \Tr e^{- \beta(E - \mu N)},
\end{equation}
where the trace operator, $\Tr$, is defined as 
\begin{equation*}
\Tr x = \sum_{\textrm{all states}} x = \sum_{n_1=0}^1 \sum_{n_2=0}^1 \dots \sum_{n_M=0}^1 x.
\end{equation*}
From the grand canonical partition function we can find the grand potential 
\begin{equation}\label{eq:grandpot}
\Omega = -k_BT \ln \Xi,
\end{equation}
in an analogous manner to which the Helmholtz free energy is obtained in the canonical ensemble [cf.\ Eq.~\eqref{eq:helmholtz1}]. The equilibrium state corresponds to the minimum of the grand potential.

\subsection{Gibbs-Bogoliubov Inequality}
We now derive and then use the Gibbs-Bogoliubov inequality to show that there exists an upper bound on the free energy and finding the minimum of this bound gives an approximation to the true free energy.

Eq.~\eqref{eq:grandpot} can be rearranged and equated to Eq.~\eqref{eq:gpartition} to give
\begin{equation}\label{eq:equating}
e^{- \beta \Omega} = \Tr e^{- \beta( E - \mu N)}.
\end{equation}
The energy of a particular state $E$ can be rewritten as
\begin{equation}\label{eq:splitenergy}
E = E_0 + E - E_0 = E_0 + \Delta E,
\end{equation}
where $E_0$ is the energy of a reference system which we choose so as to be able to evaluate the partition function. We choose the system with $\epsilon_{ij}\equiv 0$ and $V_i \neq 0$. From Eq.~\eqref{eq:equating} and \eqref{eq:splitenergy} we obtain
\begin{equation}\label{eq:bbov1}
e^{- \beta \Omega} = \Tr e^{- \beta(E_0 - \mu N)} e^{ - \beta \Delta E}.
\end{equation}
The statistical average value of any quantity $x$ in the reference system is
\begin{equation*}
\langle x \rangle_0 = \Tr \left(\frac{e^{-\beta(E_0-\mu N)}}{\Xi_0}\ x\right),
\end{equation*}
since $P_0=e^{- \beta(E_0 - \mu N)}/ \Xi_0$  [see Eq.~\eqref{eq:gprob}]. So, from \eqref{eq:bbov1} we obtain
\begin{equation}\label{eq:bbov2}
e^{-\beta \Omega} = e^{-\beta \Omega_0} \langle e^{-\beta \Delta E}\rangle_0,
\end{equation}
with $\Xi_0 =e^{-\beta \Omega_0}$ given by Eqs.~\eqref{eq:gpartition} and \eqref{eq:grandpot}. Now, since $e^{-x}$ is a convex function of $x$, then $\langle e^{-x}\rangle \geq e^{-\langle x \rangle}$ and from Eq.~\eqref{eq:bbov2} we obtain the inequality 
\begin{equation}
e^{-\beta \Omega} \geq e^{-\beta \Omega_0} e^{-\beta \langle \Delta E\rangle_0}.
\end{equation}
Taking the logarithm of this gives the Gibbs-Bogoliubov inequality\cite{Hansen06}
\begin{equation}\label{eq:bbovFinal}
\Omega \leq \Omega_0 + \langle \Delta E \rangle_0.
\end{equation}
This shows that there is an upper bound to the true grand potential $\Omega$ that depends solely on the properties of the reference system, and, more importantly, it allows us to find a `best' approximation for $\Omega$ by minimising the right hand side of the inequality. We choose $E_0$ to depend upon parameters that may be varied and perform the minimisation with respect to variations in these parameters.

To proceed, we must define $E_0$. We choose
\begin{equation}\label{eq:e0def}
E_0 = \sum_{i=1}^M (V_i +\phi_i)n_i,
\end{equation}
where $V_i$ is the external potential, and $\phi_i$ are the parameters mentioned above, which are yet to be determined. Physically, they are the (mean field) additional effective potentials that incorporate the effect of the interactions between the particles. 

The density at a particular lattice site is given by Eq.~\eqref{eq:rho_defn}. Our (mean field) approximation for this quantity is
\begin{widetext}
\begin{align}
\rho_i = \langle n_i \rangle_0 = & \Tr \left( \frac{e^{-\beta E_0- \mu N}}{Z_0} n_i \right), \nonumber\\
	= & \frac{1}{Z_0} \left[\sum_{n_1=0}^1e^{-\beta (V_1+\phi_1- \mu)n_1}\right]\dots\left[\sum_{n_i=0}^1n_ie^{-\beta (V_i+\phi_i- \mu)n_i}\right]\dots\left[\sum_{n_M=0}^1e^{-\beta (V_M+\phi_M- \mu)n_M}\right], \nonumber \\
	=& \left[ \frac{\sum_{n_1=0}^1e^{-\beta (V_1+\phi_1- \mu)n_1}}{\sum_{n_1=0}^1e^{-\beta (V_1+\phi_1- \mu)n_1}}\right]\dots \left[ \frac{\sum_{n_i=0}^1n_ie^{-\beta (V_i+\phi_i-\mu)n_i}}{\sum_{n_i=0}^1e^{-\beta (V_i+\phi_i-\mu)n_i}}\right] \dots \left[ \frac{\sum_{n_M=0}^1e^{-\beta (V_M+\phi_M-\mu)n_M}}{\sum_{n_M=0}^1e^{-\beta (V_M+\phi_M-\mu)n_M}}\right], \nonumber \\
	= & \frac{\sum_{n_i=0}^1n_ie^{-\beta (V_i+\phi_i-\mu)n_i}}{\sum_{n_i=0}^1e^{-\beta (V_i+\phi_i-\mu)n_i}}
\hspace{1cm}	=  \frac{e^{-\beta(V_i+\phi_i-\mu)}}{1+e^{-\beta(V_i+\phi_i-\mu)}}.\label{eq:averageni}
\end{align}
\end{widetext}
Also, the reference system partition function is [cf.\ Eq.~\eqref{eq:gpartition}]:
\begin{align*}
\Xi_0 = & \Tr e^{-\beta(E_0-\mu N)}, \\
	= & \Tr e^{-\beta \sum_{i=1}^M(V_i+\phi_i-\mu)n_i}, \\
	= & \prod_{i=1}^M (1+ e^{-\beta(V_i+\phi_i-\mu)}).
\end{align*}
This may then be substituted into Eq.~\eqref{eq:grandpot} to obtain the following expression for the grand potential
\begin{equation}\label{eq:omeg0}
\Omega_0 = -k_BT \sum_{i=1}^M \ln (1+ e^{-\beta(V_i+\phi_i-\mu)}).
\end{equation}
Rearranging \eqref{eq:averageni} to give ${1- \rho_i=(1+e^{- \beta(V_i+ \phi_i - \mu)})^{-1}}$ and inserting it into \eqref{eq:omeg0} gives
\begin{equation}
\Omega_0 = k_BT \sum_{i=1}^M \ln(1-\rho_i).
\end{equation}
By rewriting this as 
\begin{equation}
\Omega_0 =k_BT \sum_{i=1}^M ( \rho_i + 1 - \rho_i) \ln(1 - \rho_i),
\end{equation}
we can use Eq.~\eqref{eq:averageni} to express $ \Omega_0$ in the following form
\begin{align}\label{eq:omega0def}
\Omega_0 = k_BT & \sum_{i=1}^M \left[ \rho_i \ln \rho_i + (1- \rho_i) \ln( 1-\rho_i) \right] \nonumber \\ 
	 + & \sum_{i=1}^M \left(V_i+ \phi_i - \mu\right) \rho_i.
\end{align}
Note that when $V_i= \phi_i=0$, which corresponds to the case of a uniform fluid with $ \epsilon_{ij}=0$, this reduces to the result we saw earlier in Eq.~\eqref{eq:helmholtz3}, since $ \Omega = F - \mu N$.\cite{Hansen06,Chandler87} Returning to the general case $ \epsilon_{ij} \neq 0$, from the definition of $E_0$ in Eq.~\eqref{eq:e0def}, we find that $ \Delta E = E-E_0$ is
\begin{equation}\label{eq:deltaens}
\Delta E = -\sum_{i,j} \epsilon_{ij} n_i n_j - \sum_{i=1}^M \phi_i n_i.
\end{equation}
From Eq.~\eqref{eq:averageni}, that $ \rho_i = \langle n_i\rangle_0$, with \eqref{eq:deltaens} this gives
\begin{equation}\label{eq:deltaedef}
\langle \Delta E \rangle_0 = -\sum_{i,j} \epsilon_{ij} \rho_i \rho_j - \sum_{i=1}^M \phi_i \rho_i,
\end{equation}
where, because our reference system is non-interacting, we find that 
$$\langle n_i n_j \rangle_0 = \langle n_i\rangle_0\langle n_j\rangle_0 = \rho_i \rho_j.$$
Finally, Eqs.~\eqref{eq:omega0def} and \eqref{eq:deltaedef} can be used to obtain
\begin{align}
\hat{\Omega} & =  \Omega_0 + \langle \Delta E \rangle_0 , \nonumber\\ 
		     & 	=  k_BT  \sum_{i=1}^M \left[ \rho_i \ln \rho_i + (1- \rho_i) \ln( 1-\rho_i) \right] \nonumber \\
		     & \qquad \ -\sum_{i,j} \epsilon_{ij} \rho_i \rho_j + \sum_{i=1}^M (V_i- \mu) \rho_i. \label{eq:fullomega}
\end{align}

As discussed previously, this is an upper bound to the true grand potential $ \Omega$. One should choose the mean field $\{ \phi_i\}$ so as to minimise $\hat{ \Omega}$, in order to generate a best approximation for $ \Omega$. This is equivalent to choosing the set $\{ \rho_i\}$ so as to minimise $\hat{ \Omega}$, since the density $ \rho_i$ is defined by $ \phi_i$ [cf.\ Eq.~\eqref{eq:averageni}]. What we have done here is to derive an approximate DFT for the lattice fluid. For DFT in general, one can prove that the equilibrium fluid density profile is that which minimises the grand potential functional. \cite{Evans79}

\section{Defining the Potentials}\label{sec:potentials}
Up to this point, we have not specified the form of the potentials from the external field, or the particle interactions. We now define $ \epsilon_{ij}$ and $V_i$.

\subsection{Particle Interactions}\label{sec:IVA}

\begin{figure}[t]
\begin{center}
\begin{subfigure}{0.18\textheight}
\centering
\includegraphics{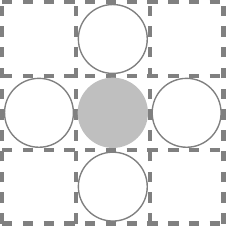}
\caption{}
\label{subfig:nearestNeighbours}
\end{subfigure}
\begin{subfigure}{0.18\textheight}
\centering
\includegraphics{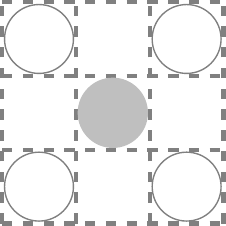}
\caption{}
\label{subfig:nextNearest}
\end{subfigure}
\caption{The distinction between, (a), the nearest neighbors (open circles) to a particle (grey circle), and, (b), the next nearest neighbors.}
\label{fig:nearest}
\end{center}
\end{figure}

The term $-\sum_{i,j} \epsilon_{ij} \rho_i \rho_j$ in Eq.~\eqref{eq:fullomega} represents the contribution to the free energy from the interactions between pairs of particles. A simple example of the continuum fluid we seek to model is made up of particles interacting via a Lennard-Jones pair potential \cite{Hansen06} of the form $v(r) = \epsilon \left[ \left(\frac{r_0}{r}\right)^{12} -2 \left(\frac{r_0}{r}\right)^6\right]$, where $r$ is the distance between pairs of particles and $r_0$ is the distance at the minimum where $v(r_0)=- \epsilon$. Given that $v(2r_0) \approx -0.03 \epsilon$, it is a good approximation to assume that each particle only interacts with the nearest and next nearest neighbouring particles, as illustrated in Fig.~\ref{fig:nearest}, and so we replace the particle interaction term in the free energy with
\begin{equation}\label{eq:internalpotential}
\sum_{i,j} \epsilon_{ij} \rho_i \rho_j \approx \epsilon_{nn} \sum_{i=1}^M \rho_i  \sum_{jnni} \rho_j + \epsilon_{nnn} \sum_{i=1}^M \rho_i \sum_{jnnni} \rho_j,
\end{equation}
where $ \epsilon_{nn}$ and $ \epsilon_{nnn}$ are the strengths of the interaction between nearest neighbour and next nearest neighbour particles, respectively. The term $\sum_{jnni} \rho_j$ denotes the sum of densities in lattice sites $j$ which are the nearest neighbours to the site $i$. Similarly, $\sum_{jnnni} \rho_j $ denotes the sum over the next nearest neighbours. We now set $ \epsilon_{nn}= \epsilon$ and $ \epsilon_{nnn}= \epsilon/4$. This ratio $ \epsilon_{nn}/ \epsilon_{nnn}$ is not the value it would have if the Lennard-Jones potential were exactly applied but it is the optimum ratio to obtain circular drops when solved in two dimensions (see \S\ref{sec:drops} below).\cite{Robbins12,Robbins11,Fomel97} Our definition captures the essence of the Lennard-Jones potential: the repulsive core is modelled by the onsite repulsion (one particle per lattice site) and the pair interaction terms crudely model the attractive forces. However, it is worth noting that even though the interaction energy between two well-separated $(r \gg r_0)$ particles can be very small, the net contribution from all such long-range interactions may be significant and neglecting them may result in the theory failing to describe some interesting physics.

\subsection{External Potential}
We assume that the interaction potential between a particle and the particles that form the wall of the container is of the Lennard-Jones form which decays for large $r$ as $v(r) \sim -r^{-6}$. Summing the potential between a single fluid particle with all of the particles in the wall yields a net potential that decays as $V(z)\sim -z^{-3}$ for $z \rightarrow \infty$, where $z$ is the perpendicular distance between the particle and the wall. We therefore assume that the wall exerts a potential of the form
\begin{equation}\label{eq:extPot}
V_i = 
\begin{cases}
\infty \quad & \textrm{if } k<1\\
-\epsilon_wk^{-3} \quad & \textrm{if } k \geq 1
\end{cases}
\end{equation}
where $ \epsilon_w$ is the parameter which defines the attractive strength of the confining wall. The integer index $k$ is the distance, in the number of lattice sites, of the particle from the wall.

\section{The Bulk Fluid Phase Diagram}\label{sec:phase}
\begin{figure}
\begin{center}
\begin{subfigure}{0.3\textheight}
\includegraphics[width=0.3\textheight]{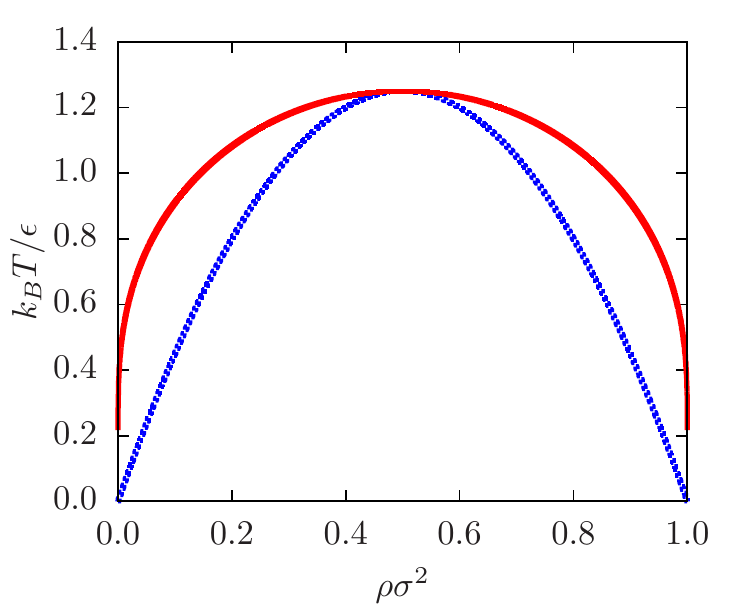}
\captionsetup{margin={0.65cm,0cm}}
\caption{}
\label{subfig:densityPhase}
\end{subfigure}
\begin{subfigure}{0.3\textheight}
\includegraphics[width=0.3\textheight]{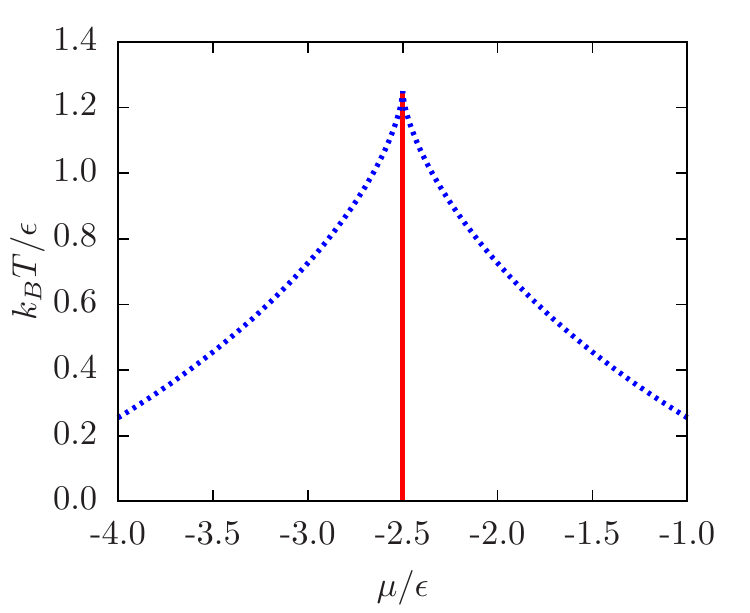}
\captionsetup{margin={0.65cm,0cm}}
\caption{}
\label{subfig:chempotPhase}
\end{subfigure}
\caption{The bulk fluid phase diagram for the 2D lattice fluid. The solid red line is the binodal and the dashed blue line is the spinodal. In (a) we display the phase diagram in the dimensionless temperature $k_BT/ \epsilon$ versus density plane and in (b) as a function of chemical potential.}
\label{fig:phase1}
\end{center}
\end{figure}

Before discussing the behaviour of the fluid at this wall, we first calculate the phase diagram of the bulk fluid, away from the influence of any interfaces. When the temperature $T$ is less than the critical temperature $T_c$, the fluid exhibits phase separation into a low density gas phase and a high density liquid phase. The binodal is the line in the phase diagram at which this transition occurs. Along the binodal, the liquid and the gas coexist in thermodynamic equilibrium, i.e.\ where the pressure, chemical potential and temperature of the liquid and gas phases are equal. The lattice gas model has a hole-particle symmetry that is not present in a continuum description, but which is useful for calculating the binodal. This symmetry arises from the fact that if we replace $n_i=1-h_i$, where $h_i$ is the hole occupation number, then the form of Eq.~\eqref{eq:energy} is unchanged. This symmetry leads to the density of the coexisting gas and liquid, $ \rho_g$ and $ \rho_l$ respectively, to be related as 
\begin{equation}\label{eq:phase1}
\rho_l= 1-\rho_g.
\end{equation}
From Eq.~\eqref{eq:fullomega} the Helmholtz free energy per lattice site, $f = F/M$, for a uniform fluid with density $ \rho$, is
\begin{equation}\label{eq:phaseextra}
f=k_BT \left[ \rho \ln \rho + (1- \rho) \ln (1- \rho)\right] - \frac{ 5 \epsilon}{2} \rho^2
\end{equation}
where $5 \epsilon/2 = \sum_{i,j} \epsilon_{ij}$ is the sum up to the next nearest neighbours interactions and includes a factor of a half to prevent double counting. The pressure in the system is\cite{Hansen06,Mandl88}
\begin{align}\label{eq:phase2}
P( \rho) & = -\left(\frac{\partial F}{\partial V}\right)_{T,N} \nonumber \\
             & =  \rho \frac{\partial f}{\partial \rho} - f \nonumber \\ 
             & = -k_BT \ln(1- \rho) - \frac{5}{2} \epsilon \rho^2.
\end{align}
The binodal curve can be found by invoking Eq.~\eqref{eq:phase1} and solving $P( \rho) = P(1- \rho)$, giving
\begin{equation}\label{eq:phase4}
\frac{k_BT}{ \epsilon} = \frac{5(2 \rho-1)}{2(\ln \rho - \ln(1- \rho))},
\end{equation}
which is displayed in Fig.~\ref{subfig:densityPhase}. The maximum on the binodal corresponds to the critical point, above which there is no gas-liquid phase separation. From the symmetry \eqref{eq:phase1}, the density at the critical point is $ \rho=1/2$ and the critical temperature is found to be $T_c= 5 \epsilon/4k_B$.

The chemical potential can also be calculated from the Helmholtz free energy as\cite{Hansen06,Mandl88}
\begin{align}\label{eq:phase5}
\mu( \rho) & = \left(\frac{\partial F}{\partial N}\right)_{T,V} \nonumber \\
                 & = \frac{\partial f}{\partial \rho}\nonumber \\
                 & = k_BT \ln \left( \frac{ \rho}{1- \rho} \right) - 5 \epsilon \rho.
\end{align}
On substituting \eqref{eq:phase4} into \eqref{eq:phase5} we find that the chemical potential at coexistence is 
\begin{equation}\label{eq:phase6}
\mu_{coex} = - \frac{ 5}{2}\epsilon,
\end{equation}
which is displayed in Fig.~\ref{subfig:chempotPhase}. The spinodal is also plotted in Fig.~\ref{fig:phase1}. The spinodal denotes the locus in the phase diagram where the compressibility is zero, i.e.\ within this curve the fluid is unstable and spontaneous phase separation occurs. The spinodal is obtained from the following condition
\begin{equation}\label{eq:phase7}
\frac{\partial^2 f}{\partial \rho^2} = 0,
\end{equation}
which, from Eq.~\eqref{eq:phaseextra}, gives the following expression for the density dependence of the temperature along the spinodal,
\begin{equation}\label{eq:phase8}
\frac{k_BT}{ \epsilon} = 5 \rho(1- \rho),
\end{equation}
also plotted in Fig.~\ref{subfig:densityPhase}. The spinodal can also be obtained as a function of $\mu$ from Eqs.\ \eqref{eq:phase5} and \eqref{eq:phase8}. The result is displayed in Fig.~\ref{subfig:chempotPhase}.

\subsection*{Exercise:}
Calculate the binodal for the case when there are only nearest neighbour interactions. What is the critical temperature?

\section{An Iterative Method for Calculating the Density Profile}\label{sec:solving}
We return now to the inhomogeneous fluid in the presence of an external potential. The equilibrium density profile is that which minimises $\hat{\Omega}$ in Eq.~\eqref{eq:fullomega}, i.e.\ it is the set $\{\rho_i\}$ which satisfiy, for all $i$,
\begin{equation*}
\frac{\partial \hat{ \Omega}}{\partial \rho_i}=0.
\end{equation*}
Performing this differentiation and rearranging gives the set of coupled equations,
\begin{equation}\label{eq:iteration}
\rho_i = (1- \rho_i) \exp\left[ \beta \left( \mu  + \epsilon \sum_{jnni} \rho_j 	+ \frac{\epsilon}{4}\sum_{jnnni} \rho_j - V_i\right) \right],
\end{equation}
which can be solved iteratively for the profile $\{\rho_i\}$. An initial approximation is required and the closer this is to the true solution, the better. We sometimes use $ \rho_i=\exp(\beta(\mu-V_i))$, which is the exact result in the low density (ideal-gas) limit or we may simply guess a likely profile. We can also use values from previous state points as an initial approximation when calculating at several state points successively, incrementing one parameter each time. With a suitable initial approximation for $\{\rho_i\}$, Eq.~\eqref{eq:iteration} can then be iterated until convergence is achieved.

It is often necessary during each iterative step to mix the result from evaluating the right hand side of Eq.~\eqref{eq:iteration}, $\rho_i^{\textrm{rhs}}$, in a linear combination with the result from the previous iteration $\rho_i^{\textrm{old}}$, i.e.
\begin{equation}\label{eq:mixing}
\rho_i^{\textrm{new}} = \alpha \rho_i^{\textrm{rhs}} + (1-\alpha) \rho_i^{\textrm{old}},
\end{equation}
where $\alpha$ may be small, typically in the range ${0.01<\alpha<0.1}$. This has the effect that only small steps are taken towards the minimum with each iteration. Omitting this mixing (i.e.\ $\alpha=1$) can give a $\rho_i^{\textrm{new}}$ that falls outside of the range (0,1) and once this happens the iterative routine breaks down.

\subsection{Normalising the Density Profile}\label{sec:VIA}
To describe an enclosed (canonical) system with fixed $N$, rather than being coupled to a reservoir which fixes $\mu$, we can think of Eq.~\eqref{eq:fullomega} as a constrained minimisation, i.e.\ as minimising the Helmholtz free energy
\begin{align}
F = k_BT & \sum_{i=1}^M \left[ \rho_i \ln \rho_i + (1- \rho_i) \ln( 1-\rho_i) \right] \nonumber \\ 
	- & \sum_{i,j} \epsilon_{ij} \rho_i \rho_j + \sum_{i=1}^M V_i \rho_i,
\end{align}
subject to the constraint that
\begin{equation}\label{eq:constraint}
N = \sum_{i=1}^M \rho_i.
\end{equation}
The chemical potential $ \mu$ is then the Lagrange multiplier. To achieve this when iteratively calculating the density profile $\{\rho_i\}$, we modify the method described above and at each iteration following \eqref{eq:mixing} the profile is renormalised: $\rho_i^{\textrm{norm}} = A\rho_i^{\textrm{new}}$, with 
\begin{equation*}
A = N \left(\sum_{i=1}^M \rho_i^{\textrm{new}}\right)^{-1},
\end{equation*}
so that the constraint \eqref{eq:constraint} is satisfied.

\subsection{Boundary Conditions}\label{sec:boundary}
At the wall, the boundary conditions (BC) for the density profile are straight forward: we simply set $\rho_i=0$ for all lattice sites $i$ `inside' the wall -- i.e.\ for $k<1$ in Eq.~\eqref{eq:extPot}. On the boundaries perpendicular to the wall, we normally use periodic BC, where it is assumed that the nearest neighbour of a lattice site on the boundary is the lattice site on the opposite boundary. For the boundary opposite the wall, periodic BC in this situation creates an artificial substrate (i.e.\ so that the fluid is confined in a capillary, between two walls). This does not cause a problem in sufficiently large systems. However, a more efficient solution is to assume that the fluid is uniform beyond the boundary opposite the wall, with specified density, e.g.\ that of the bulk gas. 

\section{Typical Solutions}\label{sec:results}
We now present results using the lattice gas model which are typical of many DFT models for a fluid exhibiting gas-liquid phase separation. After determining the equilibrium density profile using the iterative method described above we may then calculate thermodynamic quantities such as the interfacial tension or the adsorption at the wall, which is defined as
\begin{equation}\label{eq:norm3}
\Gamma=\sum_{i=1}^M( \rho_i- \rho_b),
\end{equation}
where $ \rho_b$ is the bulk density which is obtained by solving Eq.~\eqref{eq:phase5} for $\rho$. Note that $\Gamma$ is an excess number per area; the formula in Eq.~\eqref{eq:norm3} is only true when $\sigma=1$. By calculating results in the grand canonical ensemble we can track how the adsorption changes with $\mu$ (\S\ref{sec:adsorption}). Working in the canonical ensemble we can find drop profiles and calculate the contact angle that the liquid drop makes with a substrate (\S\ref{sec:drops}). From these results we can also determine if the liquid wets the substrate. We characterise a liquid as wetting a substrate when, at liquid-gas coexistence, a macroscopically thick layer of the liquid forms between the gas and the substrate.\cite{Evans89, Schick90, dietrich, BoRo01, BEIMR09, StVe09, Parry, BME87} Grand canonically, where particles are free to enter and leave the system, wetting is characterised by $\Gamma\to\infty$ as coexistence is approached $\mu\to\mu_{coex}^-$. Treating it canonically, the number of particles in the system is fixed $N=\sum_{i=1}^M\rho_i$ (using the normalisation discussed in \S\ref{sec:VIA}) so $\Gamma$ is fixed and we characterise wetting by the contact angle that a liquid drop makes with the substrate. In both cases, wetting only occurs when it is energetically beneficial, i.e.\ the liquid wetting the substrate is the state of least energy.

\subsection{One Dimensional Model}
\begin{figure}[t]
\begin{center}
\begin{subfigure}{0.18\textheight}
\centering
\includegraphics{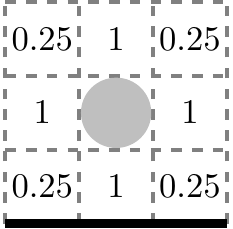}
\caption{}
\label{subfig:2d}
\end{subfigure}
\begin{subfigure}{0.18\textheight}
\centering
\includegraphics{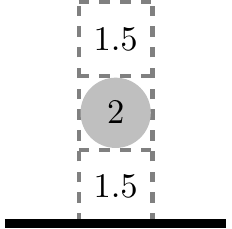}
\caption{}
\label{subfig:1d}
\end{subfigure}
\caption{Illustration of the mapping of the full 2D particle pair interactions (a) onto an effective 1D system (b). The numbers represent the contribution towards the potential (in units of $\epsilon$) from that particular lattice site with reference to the shaded particle in the centre. The 2D case on the left is that discussed above in \S\ref{sec:IVA} and on the right we display the resulting effective potential after mapping this system to 1D.}
\label{fig:2d-1d}
\end{center}
\end{figure}

So far, we have assumed for simplicity that the fluid is in 2D. However, since the density profile is defined as an average over all possible configurations [c.f.\ Eq.~\eqref{eq:rho_defn}], then if the external potential only varies in one direction [such as the potential in Eq.~\eqref{eq:extPot}], then so must the density profile. This is, of course, also the case for the 3D fluid. The equilibrium density profile must have the same symmetry as the external potential and so we may reduce the DFT equations to be solved \eqref{eq:iteration} to a one-dimensional (1D) system, consisting of a line of lattice sites extending perpendicularly away from the wall. We do this by summing over the interactions in the (transverse) direction in which the density does not vary, as illustrated in Fig.\ \ref{fig:2d-1d}. This maps the 2D system onto an effective 1D system with renormalised interactions between lattice sites and also introduces an effective on-site interaction. A similar mapping can also be done for the 3D fluid.

\subsection*{Exercise:}
(i) Implement the procedure described in \S\ref{sec:solving} on a computer for calculating the density profiles for this effective 1D model. (ii) Modify your computer code to solve for the density profile in 2D. (iii) Compare results from the two. Are they the same?

\subsection{Adsorption at the wall}\label{sec:adsorption}
\begin{figure}
\begin{center}
\begin{subfigure}[!]{0.3\textheight}
\includegraphics[width=0.3\textheight]{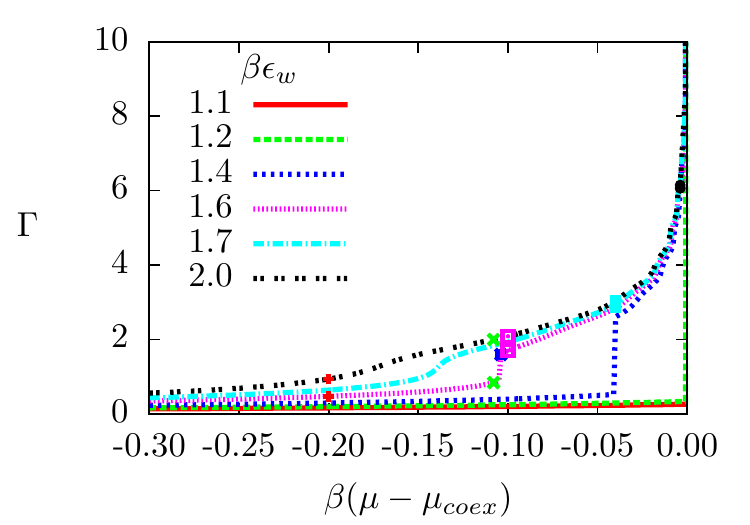}
\caption{}
\label{subfig:adsorb1}
\end{subfigure}
\begin{subfigure}[!]{0.3\textheight}
\includegraphics[width=0.3\textheight]{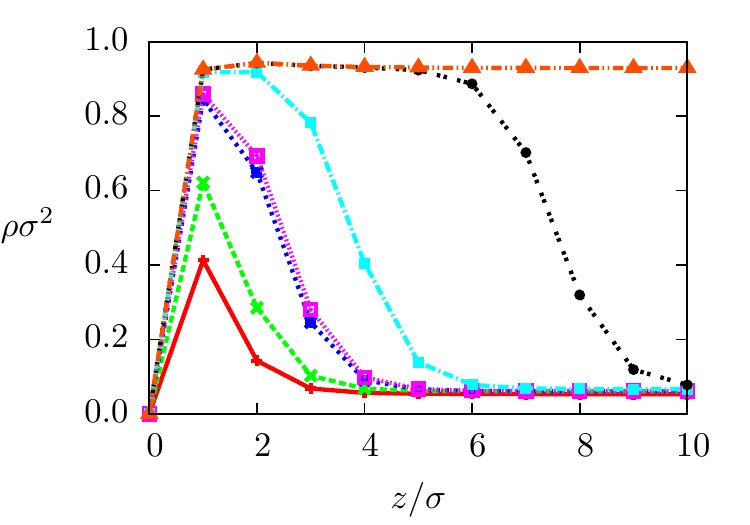}
\caption{}
\label{subfig:adsorb2}
\end{subfigure}
\begin{subfigure}[!]{0.3\textheight}
\includegraphics[width=0.3\textheight]{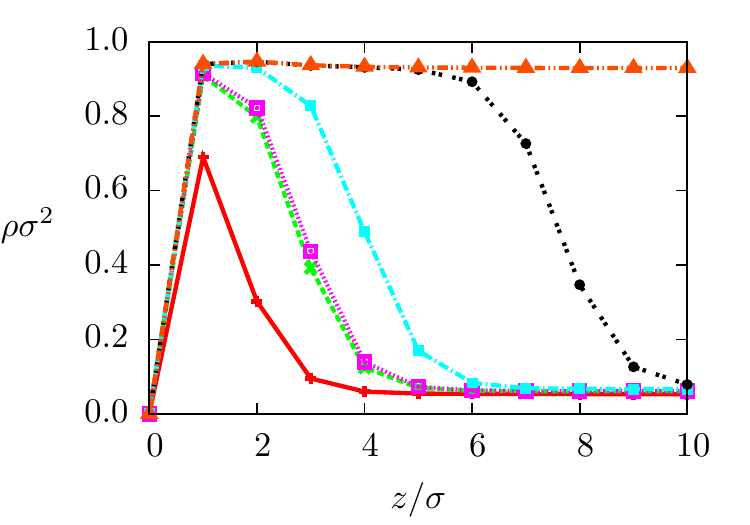}
\caption{}
\label{subfig:adsorb3}
\end{subfigure}
\caption{(a) The adsorption at the wall as the chemical potential $\mu\to\mu_{coex}^-$ for various different values of the wall attraction strength parameter $\epsilon_w$, as given in the key, for $\beta\epsilon=1.2$. In (b) we display some of the corresponding density profiles for $ \beta \epsilon_w = 1.6$, at $\beta(\mu-\mu_{coex})= -0.2$, -0.108, -0.104, -0.1, -0.04, -0.004 and 0 and in (c) we display density profiles for $ \beta \epsilon_w = 2$, at $\beta(\mu-\mu_{coex})= -0.2$, -0.108, -0.1, -0.04, -0.004 and 0. The points in (a) denote state corresponding to the profiles in (b) and (c), with matching styles and colors (color online).}
\label{fig:adsorptions}
\end{center}
\end{figure}

In Fig.~\ref{fig:adsorptions}(a) we illustrate how $\Gamma$, the adsorption at the wall, changes as the chemical potential is increased $\mu\to\mu_{coex}^-$, to approach the coexistence value in \eqref{eq:phase6}, from below. When $\mu<\mu_{coex}$ the bulk phase (away from the wall) is the gas phase, but for a wall to which the particles are attracted, the density at the substrate can be higher. As $\mu\to\mu_{coex}^-$, the adsorption increases, either diverging $\Gamma\to\infty$, when the liquid wets the wall, or remaining finite, when the liquid does not wet the wall. As $T$ or $\epsilon_w$ are changed, there is often a phase transition from one regime to the other, termed the `wetting transition'.\cite{Evans89, Schick90, dietrich, BoRo01, BEIMR09, StVe09, Parry, BME87}

The adsorption results in Fig.~\ref{fig:adsorptions}(a) are calculated for fixed $\beta\epsilon=1.2$. When the strength of the attraction due to the wall is weak, $\beta\epsilon_w<1.2$, the liquid does not wet the wall and the adsorption remains finite at coexistence, $\mu=\mu_{coex}$. However, for stronger attraction, $\beta\epsilon_w>1.2$, the wetting film thickness diverges as $\mu\to\mu_{coex}^-$. To compute these results a value of $\mu$ is set and the equilibrium profile $\{\rho_i\}$ is found. The value of $\mu$ is then incremented and the previous equilibrium solution used as the initial approximation for the next solution. At each state point the adsorption is calculated via Eq.~\eqref{eq:norm3}.

An interesting thing to note is that for some values of $\epsilon_w$, the adsorption diverges continuously (see e.g.\ the case for $\beta\epsilon_w=2$), but for other values there is a discontinuous jump in $\Gamma$. This jump is a result of crossing the `pre-wetting line'.\cite{Evans89, Schick90, dietrich, BoRo01, BEIMR09, StVe09, Parry, BME87} We see the beginings of this jump as a continuous `shoulder' for $ \beta \epsilon_w=1.7$. As $ \beta \epsilon_w$ decreases the jump becomes larger and occurs closer to $ \mu = \mu_{coex}$. The adsorption for $ \beta \epsilon_w=1.2$ remains very small until almost exactly at $ \mu= \mu_{coex}$ where it jumps to a large value. We also observe some smaller discontinuous changes in $\Gamma$ occurring after the main pre-wetting jump. These smaller jumps are `layering transitions' and are due to an additional layer of particles being discontinuously added to the adsorbed liquid film. Whilst layering transitions are observed in more sophisticated DFT theories, the underlying lattice in the present model leads to an unrealistic amplification of this effect. Figs.\ \ref{subfig:adsorb2} and \ref{subfig:adsorb3} illustrate how the density profile changes as $\mu\to\mu_{coex}^-$, for values of $\epsilon_w$ that lead to wetting of the wall. We see a layer of the liquid phase appearing against the wall, increasing in thickness as coexistence is approached. In Fig.\ \ref{subfig:adsorb2} we also see how the density profiles change discontinuously as the pre-wetting line is crossed.

Tracking the adsorption is useful for understanding how the fluid behaves as coexistence is approached. However, it ought not be used as the sole indicator of the wetting behaviour. One should also calculate the grand potential $\Omega$. It can often arise that a given density profile actually corresponds only to a local minimum of $\Omega$, but in fact the global minimum corresponds to a different density profile (e.g.\ with higher adsorption).

\subsection*{Exercise:}
Set $\beta\epsilon_w=1.3$ and calculate the density profile at coexistence $\mu=\mu_{coex}$, for a range of different `temperatures', $\beta\epsilon$. What do you find?

\subsection{Drop Profiles and Surface Tensions}\label{sec:drops}
We now return to the full 2D model and show typical density profiles corresponding to drops of liquid on a surface acting with the potential in Eq.~\eqref{eq:extPot}. We treat the system canonically, i.e.\ we normalise the system as discussed in \S\ref{sec:VIA}. We also break translational symmetry, placing the centre of mass at the horizontal midpoint.\cite{ReguerraReiss,ArcherMalijevsky}

The initial approximation for initiating our iterative procedure consists of setting the density $\rho_i=\rho_g$ everywhere, apart from in a region in the middle of the system next to the wall, where we set $\rho_i=\rho_l$. The size of this portion defines the size of the final liquid drop. The boundary conditions are as described in \S\ref{sec:boundary}, with the wall at the bottom boundary, the left and right hand sides of the lattice having periodic boundary conditions and we fix $\rho_i=\rho_g$ along the top boundary.

\begin{figure}
\begin{center}
\includegraphics[width=0.32\textheight]{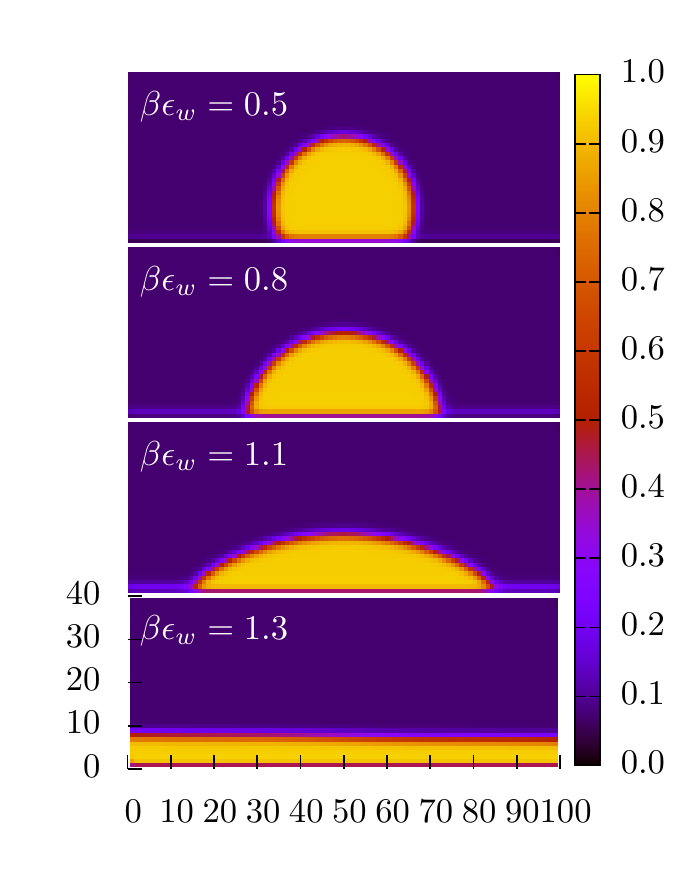}
\caption{Drop density profiles for fixed $ \beta \epsilon=1.2$ with varying values of $ \beta \epsilon_w$. The drops spread out with increasing $ \beta \epsilon_w$ until a flat film forms beyond the wetting transition.}
\label{fig:profiles}
\end{center}
\end{figure}

In Fig.~\ref{fig:profiles} we display some typical density profiles for various values of $ \beta\epsilon_w$, calculated on a $100 \times 40$ lattice, for the fluid with temperature $\beta\epsilon=1.2$. The adsorption (i.e.\ particle number) is the same in each. The liquid drop spreads out more on the substrate with the larger value of $\epsilon_w$. The contact angle $\theta$ that the drop makes with the substrate, decreases as $\epsilon_w$ is increased, so that the drop becomes broader, until complete wetting occurs at $\beta \epsilon_w \approx1.2$, when the drop becomes a flat film. 

The interfacial tension (or `surface tension' in 3D) is the excess free energy due to the presence of an interface between two phase. In the present system there are three phases: the solid (wall), liquid and gas. Thus, there are three different interfacial tensions, for the wall-liquid, wall-gas and liquid-gas interfaces, $\gamma_{wl}$, $\gamma_{wg}$ and $\gamma_{lg}$, respectively. For just the liquid and gas together, the interfacial tension $\gamma_{lg}$ leads to a liquid drop surrounded by the gas to form a circle (in 2D, or a sphere in 3D), because this shape minimises the interfacial area and therefore its contribution to the free energy. When the wall is present, which can not change, the gas and liquid must arrange themselves so as to minimise the free energy. The resulting configuration depends on the values of the interfacial tensions. The equilibrium value of the contact angle is given by Young's Equation,\cite{Evans89, Schick90, dietrich, BoRo01, BEIMR09, StVe09}
\begin{equation}\label{eq:young}
\gamma_{lg} \cos\theta=\gamma_{wg}-\gamma_{wl},
\end{equation}
which can be understood by considering the balance of the forces due to the interfacial tensions, at the point where the three phases meet.

Within the present microscopic theory, we can calculate the interfacial tensions, enabling a comparison with the macroscopic arguments that lead to to Eq.~\eqref{eq:young}. To determine $\gamma_{lg}$,  calculate the density profile through the interface between a semi-infinite slab of the liquid that is adjacent to a semi-infinite slab of the gas. This is obtained in the same manner as the density profiles at the wall in \S\ref{sec:adsorption}, but in this case we remove the wall (setting $V_i=0$ for all $i$) and set the boundary condition that $\rho_i=\rho_l$. At the other end, $\rho_i=\rho_g$, as before. The initial guess for the density profile consists of setting $\rho_i=\rho_l$ in one half of the system and $\rho_i=\rho_g$ in the other half and, of course, we must set $\mu=\mu_{coex}$. From the resulting profile $\{\rho_i\}$ we then calculate the free energy $\Omega$ from Eq.~\eqref{eq:fullomega}. The free energy without the interface (i.e.\ either full of gas or full of just the liquid) is
\begin{equation}
\Omega_0 = -pV,
\end{equation}
where $p$ is the pressure and $V$ is the volume (system size). The interfacial tension is then
\begin{equation}
\gamma_{lg}=\frac{\Omega - \Omega_0}{A}
\end{equation}
where $A$ is the length of the 2D interface. The wall-gas and wall-liquid interfacial tensions are calculated in a similar manner except we retain the wall potential and we initialise the system entirely with either the gas or the liquid density, respectively. Note that above we have solely discussed the interfacial tensions for a straight interfaces. For curved interfaces, the tensions depend on the curvature and the calculations become more involved. A discussion on some of the key issues can be found in Ref.~\onlinecite{stewart05} and references therein.

\begin{table}[t]
   \centering
   
   \begin{tabular}{cccc}
       $\beta \epsilon_w$ & $\sigma\beta\gamma_{wl}$  & $\sigma\beta\gamma_{wg}$ & $\theta$\\
      \hline
       0.5 & 0.12 & -0.05 &  115$^\circ$ \\
       0.8 & -0.16 & -0.09 & 79$^\circ$ \\
       1.0 & -0.36 & -0.13 & 52$^\circ$ \\
       1.3 & -0.68 & -0.23 & 0$^\circ$ \\
      \hline
   \end{tabular}
   \caption{Interfacial tensions and contact angle $\theta$ from Eq.~\eqref{eq:young}, for different values of the wall attraction strength $\epsilon_w$.}
   \label{tab:tensions}
\end{table}

When $\beta\epsilon=1.2$, the gas-liquid interfacial tension $\gamma_{lg}=0.38k_BT/\sigma$, corresponding to the case for the profiles displayed in Fig.~\ref{fig:profiles}. The other interfacial tensions are given in Table \ref{tab:tensions}, together with the resulting contact angle, from Eq.~\eqref{eq:young}. These are in good agreement with the contact angle one can observe from the density profiles in Fig.~\ref{fig:profiles}. However, these profiles have a diffuse interface, so there is always some uncertainty in the location of the contact line. As $\epsilon_w$ is increased the drop spreads because it is energetically beneficial to do so. Complete spreading (wetting) only occurs when the sum $\gamma_{lg}+\gamma_{wl}<\gamma_{wg}$.

\subsection*{Exercise:}
Calculate one of the density profiles from Fig.\ \ref{fig:profiles} implementing the normalisation procedure introduced in \S\ref{sec:VIA} and then plot the density contour $\rho=(\rho_g+\rho_l)/2$, that corresponds to the mid point of the liquid-gas interface. Where on this curve does the contact angle agrees with the macroscopic result in Eq.~\eqref{eq:young}? Is it where you would expect?

\section{Conclusions}\label{sec:conclusions}
We have presented a derivation of a simple lattice gas model DFT and discussed typical applications. Working with this model gives a good hands-on introduction to many of the important ideas behind DFT and gives a platform to learn about different aspects of inhomogeneous fluids such as phase diagrams, adsorption, wetting and surface tensions. Studying this `toy-model' gives students good insight and a feeling for the physics of inhomogeneous liquids, leaving them in a good position to go on and study the `real thing'.\cite{Henderson92, Rowlinson02, Davis96, Hansen06, Evans79, Lutsko10, Wu07, Wu06, Tarazona08, Lowen10}

\section*{Acknowledgements}
APH acknowledges support through a Loughborough University Graduate School Studentship. AJA thanks all the students who have done projects with him modelling inhomogeneous liquids with this lattice-gas DFT or variants of it. This paper is largely based on informal lectures and many discussions with Blesson Chacko, William Dewey, Mark Robbins and Sen Tian.


\begin{thebibliography}{1}
\bibitem{deGennes04}P.-G. de Gennes, F. Brochard-Wyart and D. Quer\'e, \textit{Capillarity and Wetting Phenomena: Drops, Bubbles, Pearls, Waves} (Springer, 2004).

\bibitem{Henderson92}D. Henderson (Ed.), \textit{Fundamentals of Inhomogeneous Fluids} (CRC Press, 1992).

\bibitem{Rowlinson02}J. S. Rowlinson, B. Widom, \textit{Molecular Theory of Capillarity} (DOVER PUBN, 2002).

\bibitem{Davis96}H. Ted Davis, \textit{Statistical Mechanics of Phases, Interfaces, and Thin Films} (Wiley-VCH, 1996).

\bibitem{Hansen06}J.-P. Hansen and I. R. McDonald, \textit{Theory of Simple Liquids}, Third Edition (Elsevier 2006).

\bibitem{Evans79}R. Evans, ``The nature of the liquid-vapour interface and other topics in the statistical mechanics of non-uniform, classical fluids'' Adv. Phys. \textbf{28} (2), 143--200 (1979).

\bibitem{Evans92}R. Evans, ``Density Functionals in the Theory of Nonuniform Fluids'' in \textit{Fundamentals of Inhomogeneous Fluids}, 85--176, D. Henderson ed. (CRC Press, 1992).

\bibitem{Lutsko10}J. F. Lutsko, ``Recent Developments in Classical Density Functional Theory'', Adv. Chem. Phys. \textbf{144}, 1--92 (2010).

\bibitem{Wu07}J. Wu and Z. Li, ``Density-Functional Theory for Complex Fluids'', Annu. Rev.Phys. Chem. \textbf{58}, 85--112 (2007).

\bibitem{Wu06}J. Wu, ``Density Functional Theory for Chemical Engineering: From Capillarity to Soft Materials'', AIChE J. \textbf{52} (3) 1169--1193 (2006).

\bibitem{Tarazona08}P. Tarazona, J. A. Cuesta and Y. Mart\'inez-Rat\'on ``Density Functional Theories of Hard Particle Systems'' , Lect. Notes Phys. \textbf{753} 247--341 (2008).

\bibitem{Lowen10}H. L\"owen, ``Density Functional Theory for Inhomogeneous Fluids II (Freezing, Dynamics, Liquid Crystals)'', Lecture Notes, 3rd Warsaw School of Statistical Physics, Warsaw University Press, 87--121 (2010).

\bibitem{endnote1} The undergraduate students involved in these projects have generally been in the final year of either a 3-year bachelors degree or a 4-year masters degree, in either Maths \& Physics or straight Maths. However, twice these were given as summer projects for very good students at an earlier stage in their studies, which worked well too. The final-year projects are typically supposed to be around 200 or 300 hours work over the academic year, including meeting with the supervisor for roughly 1 hour per week. In order to get students started and introduce to them the relevant maths and physics for these projects, 2-3 hours of informal introductory lectures are given; sections I-V of this paper are based on these. Some students are also given a computer code written in Maple that implements the method described in Sec.~VI, for the fluid in 1D with just nearest neighbour interactions, that the student then modifies to tackle their particular problem.

\bibitem{Plischke06}M. Plischke and B. Bergersen, \textit{Equilibrium Statistical Mechanics}, Second Edition (World Scientific, 2006).

\bibitem{Reichl09}L. E. Reichl, \textit{A Modern Course in Statistical Physics}, Third Edition (Wiley, 2009).

\bibitem{Robbins12}M. J. Robbins, ``Describing colloidal soft matter systems with microscopic continuum models'', PhD Thesis, Loughborough University, (2012).

\bibitem{Robbins11}M. J. Robbins, A. J. Archer and U. Thiele, ``Modelling the evaporation of thin films of colloidal suspensions using Dynamical Density Functional Theory'', J. Phys.: Condens. Matter \textbf{23} 415102 (2011).

\bibitem{Fomel97}S. Fomel and J. F. Claerbout, ``Exploring three-dimensional implicit wavefield extrapolation with the helix transform'', SEP report, \textbf{95} 43--60 (1997).

\bibitem{Chandler87}D. Chandler, \textit{Introduction to Modern Statistical Mechanics} (Oxford University Press, 1987).

\bibitem{Mandl88} F. Mandl, \textit{Statistical Physics} 2nd edition, (John Wiley \& Sons, 1988).

\bibitem{Evans89} R. Evans, ``Micoroscopic theories of simple fluids and their interfaces'' in ``Liquids at interfaces'', Les Houches, Session XLVIII 1988, Ed. J. Charvolin, J.F. Joanney and J. Zinn-Justin (Elsevier 1989).

\bibitem{Schick90}M. Schick, ``Introduction to Wetting Phenomena'' in \textit{Liquids at Interfaces, Proceedings of the Les Houches 1988 Session XLVIII}, 416--497, J. Charvolin, J.F Joanny and J. Zinn-Justin eds. (Elsevier 1990).

\bibitem{dietrich}S. Dietrich, ``Wetting phenomena'', in: Phase Transition and Critical Phenomena, vol.12, C. Domb and J.L. Lebowitz (Eds.), Academic Press, London (1988) pp. 2-218.

\bibitem{BoRo01}D. Bonn and D. Ross, ``Wetting transitions'', Rep. Prog. Phys. {\bf 64}, 1085--1163 (2001).

\bibitem{BEIMR09}D. Bonn, J. Eggers, J. Indekeu, J. Meunier and E. Rolley, ``Wetting and spreading'', Rev. Mod. Phys. {\bf 81}, 739--805 (2009).

\bibitem{StVe09}V. M. Starov and M. G. Velarde, ``Surface forces and wetting phenomena'', J. Phys.: Condens. Matter {\bf 21}, 464121 (2009).

\bibitem{Parry}A.O. Parry, ``Three-dimensional wetting revisited'', J. Phys.: Condens. Matter {\bf 8}, 10761--10778 (1996).

\bibitem{BME87}E. Bruno, U.M.B. Marconi and R. Evans, ``Phase transitions in a confined lattice gas: prewetting and capillary condensation'', Physica {\bf 141A} 187--210 (1987).



\bibitem{ReguerraReiss} D. Reguera and H. Reiss, ``The role of fluctuations in both density functional and field theory of nanosystems'', J. Chem. Phys. {\bf 120}, 2558 (2004).

\bibitem{ArcherMalijevsky} A.J. Archer and A .Malijevsky, ``On the interplay between sedimentation and phase separation phenomena in two-dimensional colloidal fluids'', Mol. Phys. {\bf 109}, 1087 (2011).

\bibitem{stewart05} M. C. Stewart and R. Evans, ``Wetting and Drying at a Curved Substrate: Long-Ranged Forces'', Phys. Rev. E {\bf 71}, 011602--011615 (2005).

\end{thebibliography}
\end{document}